\documentclass[aps,prl,amsmath, twocolumn, groupedaddress, showpacs]{revtex4}
\usepackage{amstext}
\usepackage{graphicx}
\usepackage{amssymb}
\newcommand{\ket}[1]{| #1 \rangle}



\begin{document}

\preprint{}

\title{Excitation of superconducting qubits from hot non-equilibrium quasiparticles}

\author{ J. Wenner$^1$}
\author{ Yi Yin$^1$}
\altaffiliation[Present address: ]{Department of Physics, Zhejiang University, Hangzhou 310027, China}
\author{ Erik Lucero$^1$}
\altaffiliation[Present address: ]{IBM T.J. Watson Research Center, Yorktown Heights, NY 10598, USA}
\author{ R. Barends$^1$}
\author{ Yu Chen$^1$}
\author{ B. Chiaro$^1$}
\author{ J. Kelly$^1$}
\author{ M. Lenander$^1$}
\author{ Matteo Mariantoni$^{1,2}$}
\altaffiliation[Present address: ]{Institute for Quantum Computing and Department of Physics and Astronomy, University of Waterloo, Waterloo, Ontario, Canada N2L 3G1}
\author{ A. Megrant$^{1,3}$}
\author{ C. Neill$^1$}
\author{ P. J. J. O'Malley$^1$}
\author{ D. Sank$^1$}
\author{ A. Vainsencher$^1$}
\author{ H. Wang$^{1,4}$}
\author{ T. C. White$^1$}
\author{ A. N. Cleland$^{1,2}$}
\author{ John M. Martinis$^{1,2}$}
\email{martinis@physics.ucsb.edu}

\affiliation{$^1$Department of Physics, University of California, Santa Barbara, CA 93106, USA}
\affiliation{$^2$California NanoSystems Institute, University of California, Santa Barbara, CA 93106, USA}
\affiliation{$^3$Department of Materials, University of California, Santa Barbara, CA 93106, USA}
\affiliation{$^4$Department of Physics, Zhejiang University, Hangzhou 310027, China}

\begin{abstract}
Superconducting qubits probe environmental defects such as non-equilibrium quasiparticles, an important source of decoherence. We show that ``hot'' non-equilibrium quasiparticles, with energies above the superconducting gap, affect qubits differently from quasiparticles at the gap, implying qubits can probe the dynamic quasiparticle energy distribution. For hot quasiparticles, we predict a non-neligable increase in the qubit excited state probability $P_e$. By injecting hot quasiparticles into a qubit, we experimentally measure an increase of $P_e$ in semi-quantitative agreement with the model and rule out the typically assumed thermal distribution.
\end{abstract}


\maketitle

Superconducting qubits \cite{Clarke2008,Tsai2010} are excellent candidates for building a quantum computer, with recent implementations of key quantum algorithms \cite{Reed2012,Lucero2012}. They are also sensitive probes of the physics of microscopic defects which limit coherence such as individual two-level states \cite{Lisenfeld2010,Shalibo2010}, flux noise \cite{Gustavsson2011,Choi2009,Sank2012}, and non-equilibrium quasiparticles \cite{Catelani2012,Sun2011,Lenander2011,Martinis2009,Corcoles2011,Catelani2011,CatelaniPRB2011}. The sensitivity of qubits to quasiparticles, and their ability to measure both energy emission and absorption rates, enables new measurements of the non-equilibrium properties of superconductors.

Quasiparticle-induced thermal heating has been attributed as a source \cite{Johnson2012,PalaciosLaloy2009,Corcoles2011,Geerlings2012} of qubit excited state populations \cite{Corcoles2011,Geerlings2012,Reed2010,Vijay2011,Mallet2009,Riste2012,Murch2012} and excitation rates \cite{Johnson2012} in excess of thermal equilibrium values. This is supported by the observation that the qubit excited-state population was significantly lowered when the level of stray infrared radiation, and hence quasiparticle density \cite{Barends2011}, was reduced \cite{Corcoles2011}. In these experiments, the quasiparticle-induced thermal heating necessary to produce the excited state population was thought to result in effective qubit temperatures of 70-200\,mK \cite{Johnson2012,Corcoles2011,Geerlings2012,Riste2012}, even though these temperatures are comparable to the qubit energy $E_{ge}\sim300\,$mK. This violates the typical assumptions that $k_BT\ll E_{ge}$ \cite{Paik2011,Barends2011} or $(E-\Delta)\ll E_{ge}$ \cite{Paik2011,Catelani2011} for characteristic quasiparticle energies $E$, superconducting gap $\Delta$, and dilution refrigerator temperature $T\simeq20\,\textrm{mK}$. As quasiparticle energies are thus comparable to other relevant energies, the specifics of the quasiparticle energy distribution cannot be neglected.

The quasiparticle energy distribution is frequently taken to be thermal, although sometimes with a distinct  quasiparticle temperature  from that of the environment \cite{Saira2012, Knowles2012, Fisher1999, Pekola2000,Giazotto2005}. This is assumed in single electron transistors \cite{Saira2012, Knowles2012}, normal insulator superconductor junctions \cite{Fisher1999, Pekola2000, ONeil2011, Chaudhuri2012}, superconducting tunnel junctions \cite{Giazotto2005,Heslinga1993,Wilson2001}, and kinetic inductance detetectors \cite{deVisser2011,deVisser2012}. This was even assumed when considering quasiparticle dynamics \cite{Knowles2012}.

Here, we invalidate this assumption using qubit excitation due to quasiparticles, akin to power-generating noise from quasiparticle scattering and recombination in other systems \cite{deVisser2011,deVisser2012,Wilson2001,Barends2008}. We quantitatively model how high-energy quasiparticles directly excite qubits. We experimentally test this model by injecting a non-equilibrium quasiparticle population into a superconducting qubit and using the qubit to dynamically probe this population. We find semi-quantitative agreement between our model and the experimental data presented here, showing that non-equilibrium quasiparticles provide a mechanism for the spurious excitation of superconducting qubits, distinct from thermal effects. We further rule out a thermal distribution for these quasiparticles, even with a distinct quasiparticle temperature. In addition, our approach provides a new method to study the temporal dynamics of the non-equilibrium quasiparticle energy distribution and can be used to validate alternative methods \cite{Segall2004}.

\begin{figure}
\includegraphics{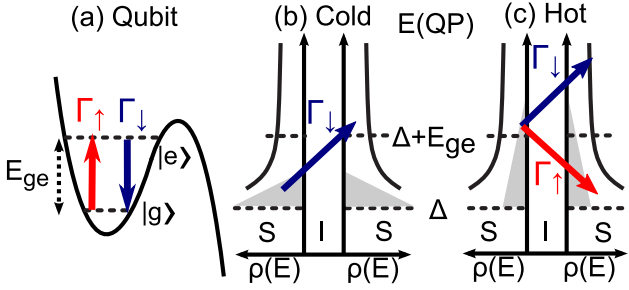}
\caption{\label{FigHotCold} (Color online) Qubit decay $\Gamma_\downarrow$ (blue) and excitation $\Gamma_\uparrow$ (red) mediated by tunneling quasiparticles. (a) Portion of the superconducting phase qubit potential energy diagram, showing the $\Gamma_\downarrow$ and $\Gamma_\uparrow$ transitions between the ground state $\ket{g}$ and the excited state $\ket{e}$, along with the qubit energy $E_{ge}$. (b) Cold non-equilibrium quasiparticles, which have energies near the superconducting gap $\Delta$, can only absorb $E_{ge}$, resulting in qubit $\Gamma_\downarrow$ decay. The density of states ($\rho(E)$, horizontal) on both sides of a Josephson junction (superconductor S - insulator I - superconductor S) is shown versus quasiparticle energy E (vertical). The quasiparticle energy distribution $f(E)$ is shown by the shaded triangles. (c) Hot non-equilibrium quasiparticles with energy above $\Delta+E_{ge}$ (portion of $f(E)$ with $E>\Delta+E_{ge}$) not only can cause qubit $\Gamma_\downarrow$ transitions but can also relax by causing qubit $\ket{g}\rightarrow\ket{e}$ transitions.}
\end{figure}

A quasiparticle tunneling through the Josephson junction barrier in a qubit can cause both excitation and dissipation in the qubit, as illustrated in Fig.\,\ref{FigHotCold}. Consider a qubit initially in its excited state: A ``cold'' quasiparticle near the gap energy can absorb the qubit transition energy $E_{ge}$ between the qubit's excited $|e\rangle$ and its ground $|g\rangle$ states, causing the qubit to switch to its ground state (blue arrows in Fig.\,\ref{FigHotCold}(a,b)). Any quasiparticle in the junction area can absorb this energy, so the qubit $|e\rangle \rightarrow |g\rangle$ decay rate $\Gamma_\downarrow$ due to this channel is proportional to the quasiparticle density $n_{qp}$.  For a qubit initially in its ground state, a ``hot'' quasiparticle sufficiently above the gap energy can excite the qubit, but only if the quasiparticle has energy greater than $\Delta + E_{ge}$ (red arrows in Fig.\,\ref{FigHotCold}(a,c)). The qubit $|g\rangle \rightarrow |e\rangle$ excitation rate $\Gamma_\uparrow$ thus depends on the energy distribution of the quasiparticle population.

If the quasiparticle population were well-described by a temperature $T \simeq 20\,\textrm{mK} \ll E_{ge}/k_B$, then a negligible qubit excitation rate $\Gamma_\uparrow$ would be expected, as in Fig.\,\ref{FigHotCold}(b).  There are however a number of processes that can produce quasiparticles with energies well above $k_B T$, which then relax via quasiparticle-phonon scattering \cite{Kaplan1976}, but for which the non-equilibrium quasiparticle occupation probability $f(E)$ still has a significant population of hot quasiparticles\cite{Martinis2009}, with energies well above $k_B T$.

\begin{figure}
\includegraphics{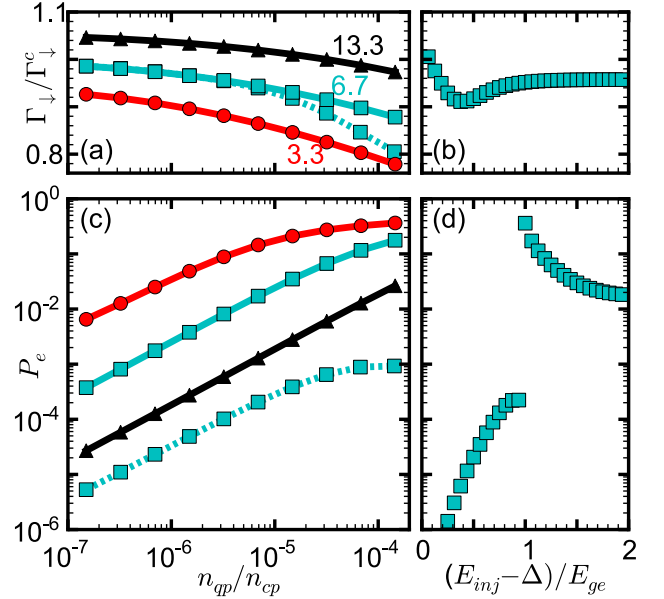}
\caption{\label{FigTheory} (Color online) Calculated effects of hot quasiparticles.  Plots are for three qubit frequencies $E_{ge}/h$: 13.3\,GHz (black), 6.7\,GHz (cyan), and 3.3\,GHz (red); lines are guides to the eye. Quasiparticle densities incorporate scattering and recombination. (a,c) Quasiparticles injected at energy $E_{inj}=1.8\Delta$ (solid lines) or $\Delta+0.94E_{ge}$ (dashed line, 6.7\,GHz), assuming $\Delta/k_B=2$\,K as for aluminum. (b,d) Quasiparticle density is $n_{qp}/n_{cp}=10^{-5}$. (a,b) Qubit decay rate $\Gamma_\downarrow$ normalized by the cold rate $\Gamma_\downarrow^c$ from Eq.\,(\ref{EqCold}), showing that $\Gamma_\downarrow^c$ is a good approximation to $\Gamma_\downarrow$. (c) Qubit $\ket{e}$ state probability $P_e=\Gamma_\uparrow/(\Gamma_\downarrow+\Gamma_\uparrow)$ vs. normalized quasiparticle density $n_{qp}/n_{cp}$. The $\ket{e}$ state occupation increases linearly with the quasiparticle density for low densities. (d) Qubit probability $P_e$ vs. injection energy $E_{inj}$ expressed as fraction of $E_{ge}$ above $\Delta$. Injecting low energy quasiparticles results in a greatly reduced $P_e$. For $E_{inj}\gtrsim\Delta+1.7E_{ge}$, $P_e$ is essentially constant.}
\end{figure}

In order to model the steady-state quasiparticle distribution, we assume quasiparticles are injected in the junction at a constant rate at an energy $E_{inj}$ well above $\Delta+E_{ge}$, with the resulting quasiparticle density $n_{qp}$ scaling as the square root of the injection rate; we verify below (discussion of Fig.\,\ref{FigTheory}(d)) that the qubit excited state probability $P_e$ is independent of the injection energy. Steady state is achieved by balancing phonon scattering and quasiparticle injection and recombination\cite{ExplainLenander}. Although the resulting steady-state occupation probability $f(E)$ has a similar dependence on quasiparticle energy as a 70\,mK thermal distribution for $\Delta<E<1.4\Delta$ \cite{ExplainAnsmann}, no effective temperature can fully describe $f(E)$ for all energies, implying a non-thermal distribution.

With $f(E)$ determined in this manner, we calculate the qubit excitation rate $\Gamma_\uparrow$ and decay rate $\Gamma_\downarrow$. For a tunnel junction with resistance $R_T$ and capacitance $C$, and for normalized quasiparticle density of states $\rho(E) = E/\sqrt{E^2-\Delta^2}$, the qubit decay (excitation) rate induced by all quasiparticles is \cite{Lenander2011,Martinis2009}
\begin{equation}
\Gamma_{\downarrow(\uparrow)} = \frac{1+\cos\phi}{R_T C}
\int_{\Delta(+E_{ge})}^\infty \frac{dE}{E_{ge}} \frac{EE_f+\Delta^2}{E E_f} \rho(E)\rho(E_f) f(E),
\label{EqHot}
\end{equation}
where the final quasiparticle energy $E_f=E+E_{ge}$ ($E_f=E-E_{ge}$) is higher (lower) than $E$ due to the absorption (emission) of the qubit energy $E_{ge}$.  Here, $\phi$ is the junction phase, which is typically $\phi\simeq0$ for the transmon and $\phi\simeq\pi/2$ for the phase qubit. For cold non-equilibrium quasiparticles, corresponding to $f(E)$ having population only at the gap energy $\Delta$, this integral gives
\begin{equation}
\Gamma_\downarrow^c = \frac{1+\cos\phi}{\sqrt{2}\ R_T C}
\Big( \frac{\Delta}{E_{ge}} \Big)^{3/2} \frac{n_{qp}}{n_{cp}},
\label{EqCold}
\end{equation}
Here, $n_{cp} = D(E_F)\Delta$ is the Cooper pair density, $n_{qp}=2D(E_F) \int_\Delta^\infty \rho(E) f(E) dE$ is the quasiparticle density, incorporating both hot and cold quasiparticles, and $D(E_F)/2$ is the single spin density of states.  This is the standard result for quasiparticle dissipation \cite{Catelani2011}, and is equivalent \cite{Lenander2011} to the Mattis-Bardeen theory \cite{Mattis1958} for $\phi = 0$.

In Fig.\,\ref{FigTheory}(a,b) we plot $\Gamma_\downarrow/\Gamma_\downarrow^c$, the ratio of the numerically-integrated qubit decay rate $\Gamma_\downarrow$ assuming both hot and cold non-equilibrium quasiparticles (Eq.\,\ref{EqHot}) to the rate $\Gamma_\downarrow^c$ for cold quasiparticles at the gap (Eq.\,\ref{EqCold}), with $f(E)$ calculated as explained above. We see that $\Gamma_\downarrow \approx \Gamma_\downarrow^c$ for a range of parameters, so the quasiparticle-induced decay rate is determined primarily by the total quasiparticle density $n_{qp}$ and depends only weakly on the quasiparticle occupation distribution.

However, the quasiparticle distribution is key to characterizing the quasiparticle-induced steady-state excited state population, $P_e=\Gamma_\uparrow/(\Gamma_\downarrow+\Gamma_\uparrow)$. Using the same $f(E)$ as before, we calculate the probabilities plotted in Fig.\,\ref{FigTheory}(c). Notice that a non-negligible probability of a few percent can be obtained for quite modest quasiparticle densities.  The probability $P_e$ decreases for smaller quasiparticle densities and for larger qubit energies, as expected.  For small occupation probabilities, this result can be approximated by the fit function $P_e \simeq 2.17(n_{qp}/n_{cp})(\Delta/E_{ge})^{3.65}$.

\begin{figure}
\includegraphics{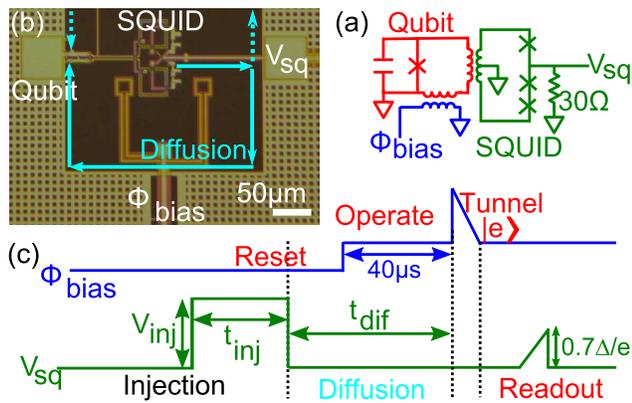}
\caption{\label{FigDiagram} (Color online) Experimental apparatus and protocol. (a) Schematic of phase qubit, with voltage $V_{sq}$ applied to readout SQUID and flux bias $\Phi_{bias}$. Quasiparticles generated with the SQUID can only reach the qubit through the common ground. (b) Photo of qubit. Two paths for quasiparticles to diffuse to the qubit junction are indicated by dotted and solid arrows. (c) Pulse sequence. Quasiparticles are generated with a voltage pulse on the SQUID line of amplitude $V_{inj}>2(\Delta+E_{ge})/e$ for a time $t_{inj}$, which are varied to adjust the quasiparticle density. After a delay time $t_{dif}$, 40$\mu$s of which is at the operating bias, allowing quasiparticles to diffuse to the qubit, the qubit state is projected with a flux bias pulse and read out using a voltage pulse on the SQUID.}
\end{figure}

To determine the sensitivity of this result to the quasiparticle injection energy, we also calculated $P_e$ as a function of the injection energy $E_{inj}$. As shown in Fig.\,\ref{FigTheory}(d), $P_e$ is independent of the injection energy for $E_{inj} \gtrsim \Delta+1.7E_{ge}$; hence, for sufficiently large injection energies, the actual injection value is unimportant.  In addition, we see that $P_e$ is significantly suppressed for quasiparticle energies below $\Delta+E_{ge}$, demonstrating that cold quasiparticles do not excite the qubit. The maximum at $E_{inj}=\Delta+E_{ge}$ is caused by the peaked final state density of states $\rho(E_f)$ at $E_f=\Delta$ in the expression for $\Gamma_\uparrow$.

\begin{figure}
\includegraphics{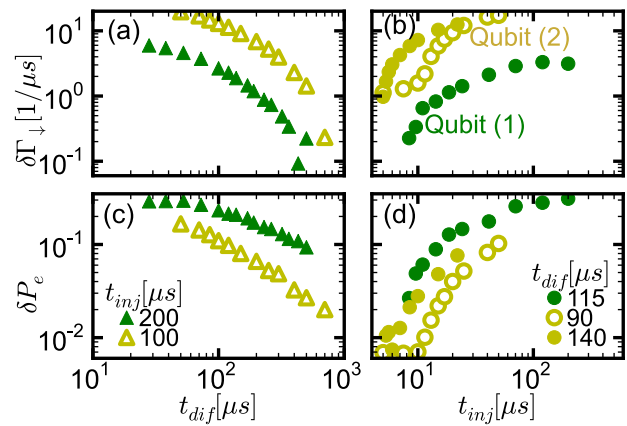}
\caption{\label{FigTimes} (Color online) Experimental impacts of injected quasiparticles. Increases in (a,b) qubit decay rate $\delta\Gamma_\downarrow$ and (c,d) qubit excited state probability $\delta P_e$ are measured with respect to values without quasiparticle injection. The quasiparticle density was varied by changing (b,d) the length $t_{inj}$ of the injection pulse or (a,c) the diffusion time $t_{dif}$. Data are for two slightly different designs: 1 (Ref.\,\cite{Lucero2012}) and 2 (Ref.\,\cite{Yin2012}); experimental parameters are in the supplementary material \cite{Supp}. These increases direcly demonstrate a non-equilibrium distribution.}
\end{figure}

To experimentally test these concepts, we performed an experiment in which we deliberately injected quasiparticles into a phase qubit (Fig.\,\ref{FigDiagram}(a,b)) and measured the excited state probability $P_e$ and the increase in the qubit decay rate $\delta\Gamma_\downarrow$, which is proportional to $n_{qp}/n_{cp}$ (Eq.\,\ref{EqCold}). As shown in Fig.\,\ref{FigDiagram}(c), we generated quasiparticles by applying a voltage pulse above the gap voltage $\Delta/e$ to the qubit's readout superconducting quantum interference device (SQUID). The duration $t_{inj}$ of this pulse was varied to adjust the quasiparticle density and $f(E)$. We reset the qubit into the $\ket{g}$ state using $\Phi_{bias}$ and then waited a variable time $t_{dif}$ following the pulse, giving the quasiparticles time to diffuse to the qubit and allowing study of the temporal dynamics. After the qubit control pulses, we measured the qubit $P_e$, reading out the qubit by increasing $V_{sq}$ to approximately $0.7\Delta/e$ to switch the SQUID into the normal state while minimizing quasiparticle generation. Note this is similar to previous work \cite{Lenander2011}, except here the qubit measurements include the enhancement $\delta P_e$ versus the energy relaxation rate increase $\delta\Gamma_\downarrow$. The increases $\delta P_e$ and $\delta\Gamma_\downarrow$, as shown in Fig.\,\ref{FigTimes}, are measured by comparing $P_e$ and $\Gamma_\downarrow$ both with and without the quasiparticle injection pulse.

\begin{figure}[t]
\includegraphics{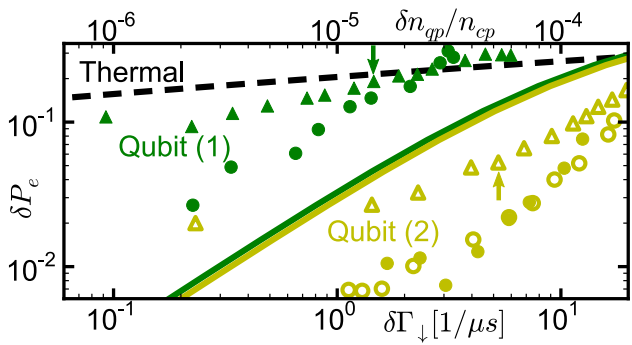}
\caption{\label{FigData} (Color online) Comparison of data and model. The experimental increase in qubit excited state probability $\delta P_e$ from Fig.\,\ref{FigTimes} is parametrically plotted vs. increases in qubit decay rate $\delta\Gamma_\downarrow$, equivalent to the quasiparticle density $\delta n_{qp}/n_{cp}$. Symbols are as in Fig.\,\ref{FigTimes}. The power law dependence of $\delta P_e$ on $\delta\Gamma_\downarrow$ is reduced to the left of $t_{inj}+t_{dif}\sim350\,\mu$s (arrows). Theoretical predictions (solid lines) are for quasiparticles injected at $1.9\Delta$ using measured $E_{ge}\simeq0.15\Delta$. Dashed line is for a Fermi-Dirac quasiparticle distribution at quasiparticle temperatures selected for $\delta\Gamma_\downarrow$.}
\end{figure}

In Fig.\,\ref{FigData}, we plot the observed changes $\delta P_e$ versus $\delta\Gamma_\downarrow$ for two different phase qubits (details in \cite{Lucero2012,Yin2012}, parameters in the supplementary material \cite{Supp}) as we varied $t_{dif}$ and $t_{inj}$ (Fig.\,\ref{FigTimes}). We find that $\delta P_e$ monotonically increases with $\delta\Gamma_\downarrow$, thus scaling with quasiparticle density, as predicted by the theory. As $\delta\Gamma_\downarrow\not=0$, the quasiparticles are in a non-equilibrium distribution. In addition, we were able to increase $P_e$ by more than 10\%, demonstrating the presence of hot quasiparticles and directly showing that hot quasiparticles can significantly excite the qubit.

The quasiparticle density was changed in three ways: Varying the diffusion time $t_{dif}$ (triangles), the injection time $t_{inj}$, and the injection voltage $V_{inj}$ (open vs. closed).  There is essentially no difference between changing the injection voltage and the injection time; this makes sense, as $P_e$ is relatively insensitive to the injection energy (Fig.\,\ref{FigTheory}(d)). In addition, for times less than $t_{inj}+t_{dif}\leq350\,\mu$s (equality denoted by arrows), the data where $t_{dif}$ was varied is similar to the data where $t_{inj}$ is varied. However, we observed a significant difference with varying the diffusion time for $t_{inj}+t_{dif}>350\,\mu$s. This is not surprising, as $P_e$ is sensitive to the quasiparticle energy distribution, which changes with the diffusion time. As shown in Fig.\,\ref{FigDiagram}(b), quasiparticles generated at the measurement SQUID must diffuse through approximately 700$\,\mu$m of ground plane metal in order to reach the qubit junction. This gives time for quasiparticle scattering and relaxation with respect to the injected distribution. In fact, this transition time is consistent with typical quasiparticle recombination times \cite{Barends2008,Mazin2010}, where $f(E)$ is expected to become dominated by thermal equilibrium excitation independent of quasiparticle density.

We also considered a thermal quasiparticle distribution, where the quasiparticle temperature was set to give the appropriate $\delta\Gamma_\downarrow$ (dashed line). The only experimental data fit by the thermal model is when $t_{dif}$ is varied for qubit (1), consistent with much of these data having $t_{inj}+t_{dif}$ longer than typical quasiparticle recombination times. For qubit (2) or when $t_{inj}$ is varied, this thermal model does not fit the experimental data in magnitude or slope, indicating a non-thermal distribution.

Predictions from the model are plotted in Fig.\,\ref{FigData} as solid lines.  Although the power-law dependence (slopes) between model and measurement are in reasonable agreement, the model $\delta P_e$ is low by about a factor of three for qubit (1) and high by about a factor of four for qubit (2).  This can not be attributed to the small differences in the qubit geometries, because a third device (not shown), with the same design as (1), gave data about a factor of two lower than the model.  These differences could be due to subtleties in the model not considered here. For instance, there can be sample-to-sample variation such as a difference in film thickness or film quality, yielding variations in the gap energy, in the quasiparticle diffusion path, and in junction parameters. In addition, we note that the calculation of $P_e$ assumes non-equilibrium quasiparticle relaxation through electron-phonon scattering and recombination. Other effects such as quasiparticle diffusion and trapping of quasiparticles from non-uniform gaps may significantly affect the distribution $f(E)$, altering the prediction for the excitation rate; however, such effects are difficult to model due to the qubit geometry \cite{Lenander2011}. We further assume a constant quasiparticle injection rate; while this describes infrared-generated quasiparticles, here it may mask effects of the repetition rate and quasiparticle-induced heating. In addition, even though quasiparticles are expected to be generated with energies of $eV_{inj}/2$, in reality there will be a distribution of quasiparticle energies centered around this value for a given $V_{inj}$, so even for $eV_{inj}>2(\Delta+E_{ge})$, some of the generated quasiparticles may instead be cold.  We conclude that the simple theoretical model used here for the quasiparticle energy distribution $f(E)$ is only semiquantitative, predicting $\Gamma_\uparrow$ in the non-thermal regime to about a factor of 4.

In conclusion, we have shown that ``hot'' quasiparticles with energies greater than $\Delta+E_{ge}$ can cause significant ground-to-excited state transitions in superconducting qubits. This is in contrast to ``cold'' quasiparticles solely at the gap, which only cause superconducting qubits to relax. This means quasiparticles cannot be adequately described by a single parameter such as $n_{qp}/n_{cp}$ or temperature. As illustrated by varying both $t_{inj}$ and $t_{dif}$, the particular quasiparticle distribution affects the observed qubit excitation probability. This theory semiquantitatively matches the observed behavior of the qubit excitation probability versus quasiparticle density, indicating one must consider a non-thermal quasiparticle distribution.

\begin{acknowledgments}
Devices were made at the UC Santa Barbara Nanofabrication Facility, a part of the NSF-funded National Nanotechnology Infrastructure Network. This research was funded by the Office of the Director of National Intelligence (ODNI), Intelligence Advanced Research Projects Activity (IARPA), through Army Research Office grant W911NF-09-1-0375. All statements of fact, opinion or conclusions contained herein are those of the authors and should not be construed as representing the official views or policies of IARPA, the ODNI, or the U.S. Government. MM acknowledges support from an Elings Postdoctoral Fellowship. RB acknowledges support from the Rubicon program of the Netherlands Organisation for Scientific Research.
\end{acknowledgments}


\begin{thebibliography}{11}
\bibitem{Clarke2008} J. Clarke and F. Wilhelm, Nature \textbf{453}, 1031 (2008).
\bibitem{Tsai2010} J.-S. Tsai, Proc. Jpn. Acad., Ser. B \textbf{86}, 275 (2010).
\bibitem{Reed2012} M. D. Reed, L. DiCarlo, S. E. Nigg, L. Sun, L. Frunzio, S. M. Girvin, and R. J. Schoelkopf, Nature \textbf{482}, 382 (2012).
\bibitem{Lucero2012} E. Lucero \textit{et al.}, Nature Phys., \textbf{8}, 719 (2012).
\bibitem{Lisenfeld2010} J. Lisenfeld, C. M\"{u}ller, J. H. Cole, P. Bushev, A. Lukashenko, A. Shnirman, and A. V. Ustinov, Phys. Rev. Lett. \textbf{105}, 230504 (2010).
\bibitem{Shalibo2010} Y. Shalibo, Y. Rofe, D. Shwa, F. Zeides, M. Neeley, J. M. Martinis, and N. Katz, Phys. Rev. Lett. \textbf{105}, 177001 (2010).
\bibitem{Gustavsson2011} S. Gustavsson, J. Bylander, F. Yan, W. D. Oliver, F. Yoshihara, and Y. Nakamura, Phys. Rev. B \textbf{84}, 014525 (2011).
\bibitem{Sank2012} D. Sank \textit{et al.}, Phys. Rev. Lett. \textbf{109}, 067001 (2012).
\bibitem{Choi2009} S. K. Choi, D.-H. Lee, S. G. Louie, and J. Clarke, Phys. Rev. Lett. \textbf{103}, 197001 (2009).
\bibitem{Catelani2012} G. Catelani, S. E. Nigg, S. M. Girvin, R. J. Schoelkopf, and L. I. Glazman, Phys. Rev. B \textbf{86}, 184514 (2012).
\bibitem{Martinis2009} J. M. Martinis, M. Ansmann, and J. Aumentado, Phys. Rev. Lett. \textbf{103}, 097002 (2009).
\bibitem{Lenander2011} M. Lenander \textit{et al.}, Phys. Rev. B \textbf{84}, 024501 (2011).
\bibitem{Catelani2011} G. Catelani, J. Koch, L. Frunzio, R. J. Schoelkopf, M. H. Devoret, and L. I. Glazman, Phys. Rev. Lett. \textbf{106}, 077002 (2011).
\bibitem{CatelaniPRB2011} G. Catelani, R. J. Schoelkopf, M. H. Devoret, and L. I. Glazman, Phys. Rev. B \textbf{84}, 064517 (2011).
\bibitem{Sun2011} L. Sun \textit{et al.}, Phys. Rev. Lett. \textbf{108}, 230509 (2012).
\bibitem{Corcoles2011} A. D. C\'{o}rcoles, J. M. Chow, J. M. Gambetta, C. Rigetti, J. R. Rozen, G. A. Keefe, M. B. Rothwell, M. B. Ketchen, and M. Steffen, Appl. Phys. Lett. \textbf{99}, 181906 (2011).
\bibitem{Geerlings2012} K. Geerlings, S. Shankar, Z. Leghtas, M. Mirrahimi, L. Frunzio, R. J. Schoelkopf, and M. H. Devoret, in \textit{Bulletin of the American Physical Society}, proceedings of the APS March Meeting, Boston, Massachusetts (American Physical Society, 2012), Z29.00009.
\bibitem{Johnson2012} J. E. Johnson, C. Macklin, D. H. Slichter, R. Vijay, E. B. Weingarten, J. Clarke, and I. Siddiqi, Phys. Rev. Lett. \textbf{109}, 050506 (2012).
\bibitem{PalaciosLaloy2009} A. Palacios-Laloy, F. Mallet, F. Nguyen, F. Ong, P. Bertet, D. Vion, and D. Esteve, Phys. Scripta \textbf{T137}, 014015 (2009).
\bibitem{Riste2012} D. Rist\`{e}, C. C. Bultink, K. W. Lehnert, and L. DiCarlo, Phys. Rev. Lett. \textbf{109}, 240502 (2012).
\bibitem{Reed2010} M. D. Reed, L. DiCarlo, B. R. Johnson, L. Sun, D. I. Schuster, L. Frunzio, and R. J. Schoelkopf, Phys. Rev. Lett. \textbf{105}, 173601 (2010).
\bibitem{Vijay2011} R. Vijay, D. H. Slichter, and I. Siddiqi, Phys. Rev. Lett. \textbf{106}, 110502 (2011).
\bibitem{Murch2012} K. W. Murch, U. Vool, D. Zhou, S. J. Weber, S. M. Girvin, and I. Siddiqi, Phys. Rev. Lett. \textbf{109}, 183602 (2012).
\bibitem{Mallet2009} F. Mallet, F. R. Ong, A. Palacios-Laloy, F. Nguyen, P. Bertet, D. Vion, and D. Esteve, Nature Phys. \textbf{5}, 791 (2009).
\bibitem{Barends2011} R. Barends \textit{et al.}, Appl. Phys. Lett. \textbf{99}, 113507 (2011).
\bibitem{Paik2011} H. Paik \textit{et al.}, Phys. Rev. Lett. \textbf{107}, 240501 (2011).
\bibitem{Saira2012} O.-P. Saira, A. Kemppinen, V. F. Maisi, and J. P. Pekola, Phys. Rev. B \textbf{85}, 012504 (2012).
\bibitem{Knowles2012} H. S. Knowles, V. F. Maisi, and J. P. Pekola, Appl. Phys. Lett. \textbf{100}, 262601 (2012).
\bibitem{Giazotto2005} F. Giazotto and J. P. Pekola, J. Appl. Phys. \textbf{97}, 023908 (2005).
\bibitem{Fisher1999} P. A. Fisher, Ph.D. thesis, Harvard University, 1999.
\bibitem{Pekola2000} J. P. Pekola, D. V. Anghel, T. I. Suppula, J. K. Suoknuuti, A. J. Manninen, and M. Manninen, Appl. Phys. Lett. \textbf{76}, 2782 (2000).
\bibitem{ONeil2011} G. C. O'Neil, P. J. Lowell, J. M. Underwood, and J. N. Ullom, Phys. Rev. B \textbf{85}, 134504 (2012).
\bibitem{Chaudhuri2012} S. Chaudhuri and I. J. Maasilta, Phys. Rev. B \textbf{85}, 014519 (2012).
\bibitem{Heslinga1993} D. R. Heslinga and T. M. Klapwijk, Phys. Rev. B \textbf{47}, 5157 (1993).
\bibitem{Wilson2001} C. M. Wilson, L. Frunzio, and D. E. Prober, Phys. Rev. Lett. \textbf{87}, 067004 (2001).
\bibitem{deVisser2011} P. J. de Visser, J. J. A. Baselmans, P. Diener, S. J. C. Yates, A. Endo, and T. M. Klapwijk, Phys. Rev. Lett. \textbf{106}, 167004 (2011).
\bibitem{deVisser2012} P. J. de Visser, J. J. A. Baselmans, S. J. C. Yates, P. Diener, A. Endo, and T. M. Klapwijk, Appl. Phys. Lett. \textbf{100}, 162601 (2012).
\bibitem{Barends2008} R. Barends, J. J. A. Baselmans, S. J. C. Yates, J. R. Gao, J. N. Hovenier, and T. M. Klapwijk, Phys. Rev. Lett. \textbf{100}, 257002 (2008).
\bibitem{Segall2004} K. Segall, C. Wilson, L. Li, L. Frunzio, S. Friedrich, M. C. Gaidis, and D. E. Prober, Phys. Rev. B \textbf{70}, 214520 (2004).
\bibitem{Kaplan1976} S. B. Kaplan, C. C. Chi, D. N. Langenberg, J. J. Chang, S. Jafarey, and D. J. Scalapino, Phys. Rev. B \textbf{14},4854 (1976).
\bibitem{ExplainLenander} Equations (C3-C7) of \cite{Lenander2011} are used to calculate the quasiparticle number distribution for a given injection rate. The quasiparticle energy distribution $f(E)$ is then calculated using Eq.\,(C1), and the total quasiparticle density is calculated with Eq.\,(C2).
\bibitem{ExplainAnsmann} This is determined from the nearly-constant slope in the region $1<E/\Delta<1.4$ of Fig.\,1 in \cite{Martinis2009} given $n_{qp}/n_{cp}=1.8\times10^{-6}$ (calculated using \cite{ExplainLenander}).
\bibitem{Mattis1958} D. Mattis and J. Bardeen, Phys. Rev. \textbf{111}, 412 (1958).
\bibitem{Yin2012} Y. Yin \textit{et al.}, Phys. Rev. Lett. \textbf{110}, 107001 (2013).
\bibitem{Supp} See Supplemental Material at [URL will be inserted by publisher] for experimental parameters and description of device.
\bibitem{Mazin2010} B. A. Mazin, D. Sank, S. McHugh, E. A. Lucero, A. Merrill, J. Gao, D. Pappas, D. Moore, and J. Zmuidzinas, Appl. Phys. Lett. \textbf{96}, 102504 (2010).

\end{thebibliography}
\end{document}


\preprint{}

\title{Supplementary Material for ``Excitation of superconducting qubits from hot non-equilibrium quasiparticles''}

\author{ J. Wenner$^1$}
\author{ Yi Yin$^1$}
\author{ Erik Lucero$^1$}
\author{ R. Barends$^1$}
\author{ Yu Chen$^1$}
\author{ B. Chiaro$^1$}
\author{ J. Kelly$^1$}
\author{ M. Lenander$^1$}
\author{ Matteo Mariantoni$^{1,2}$}
\author{ A. Megrant$^{1,3}$}
\author{ C. Neill$^1$}
\author{ P. J. J. O'Malley$^1$}
\author{ D. Sank$^1$}
\author{ A. Vainsencher$^1$}
\author{ H. Wang$^{1,4}$}
\author{ T. C. White$^1$}
\author{ A. N. Cleland$^{1,2}$}
\author{ John M. Martinis$^{1,2}$}
\email{martinis@physics.ucsb.edu}

\affiliation{$^1$Department of Physics, University of California, Santa Barbara, CA 93106, USA}
\affiliation{$^2$California NanoSystems Institute, University of California, Santa Barbara, CA 93106, USA}
\affiliation{$^3$Department of Materials, University of California, Santa Barbara, CA 93106, USA}
\affiliation{$^4$Department of Physics, Zhejiang University, Hangzhou 310027, China}

\renewcommand{\thefigure}{S\arabic{figure}} 
\renewcommand{\thetable}{S\arabic{table}} 
\renewcommand\thepage {S\arabic{page}}
\setcounter{figure}{0}
\setcounter{table}{0}
\setcounter{page}{1}

\begin{abstract}
Here, we present experimental details along with a description of the phase qubits used.
\end{abstract}

\maketitle

\begin{table}[t!]
\caption{Experimental parameters. The injection voltage $V_{inj}$ is provided assuming a junction superconducting gap $\Delta/e=2$K as for aluminum. We also include parameters measured independently without quasiparticle injection: excited state probability $P_e^0$, qubit decay rate $\Gamma_\downarrow^0$, qubit frequency $E_{ge}/h$, and qubit critical current $I_c$. Symbols are as used in Figs.\,4-5.}
\begin{tabular}{
||l||c||c|c|c|c||}
\hline
\hline
Data & $V_{inj}$ & $P_e^0$ & $1/\Gamma_\downarrow^0$ & $E_{ge}/h$ & $I_c$ \\
\hline
\hline
(1) Triangles & 0.57\,mV,3.5$\Delta/e$ & 3-4\% & 880\,ns & 5.8\,GHz & 1.0\,$\mu$A  \\
\hline
(1) Circles   & 0.57\,mV,3.5$\Delta/e$ & 1-9\% & 880\,ns & 5.5\,GHz & 1.0\,$\mu$A  \\
\hline
(2) Open      & 0.41\,mV,2.6$\Delta/e$ & 4.0\% & 380\,ns & 6.1\,GHz & 1.7\,$\mu$A  \\
\hline
(2) Closed    & 0.59\,mV,3.7$\Delta/e$ & 4.7\% & 380\,ns & 6.1\,GHz & 1.7\,$\mu$A  \\
\hline
\hline
\end{tabular}
\label{TabData}
\end{table}

\begin{figure}[t!]
\includegraphics{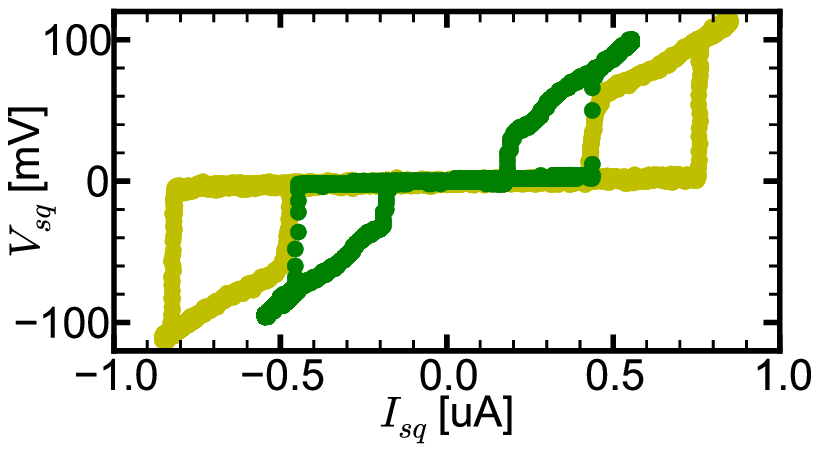}
\caption{\label{FigIV} (Color online) SQUID voltage $V_{sq}$ vs current $I_{sq}$. The SQUIDs are shunted by a $30\,\Omega$ resistor, as shown in Fig.\,3(a). The qubits are labeled as in Table \ref{TabData} and Figs.\,4-5.}
\end{figure}

Both qubits tested were fabricated using a multi-layer process with optical lithography. The qubit and SQUID junctions are Al/AlO$_x$/Al on a sapphire substrate, while the parallel-plate capacitor shunting the qubit had a hydrogenated amorphous silicon dielectric with a capacitance of 1\,pF.

In Fig.\,\ref{FigIV} and Table \ref{TabData}, we present the qubit and SQUID parameters for qubits 1 (Ref.\,\cite{Lucero2012}) and 2 (Ref.\,\cite{Yin2012}). The SQUID current vs. voltage for both devices is shown in Fig.\,\ref{FigIV}, whereas the qubit junction critical current and qubit frequency are in Table \ref{TabData}. Further, we include in Table \ref{TabData} the qubit decay rate and excited state probabilities without quasiparticle injection to which the data with quasiparticle injection is compared; the range in excited state probabilities for Qubit (1) indicates different no-injection values correspond to different data points.

In Figs.\,4-5, we only present data where the second excited state is not appreciably populated. To do this, we use the measurement fidelities for the ground and first and second excited states as measured without quasiparticle injection to correct the probabilities of being in these states with quasiparticle injection \cite{Bialczak2010}. We then compare the resulting quasiparticle-induced excited state probability with the raw increase in the excited state probability due to quasiparticle injection and exclude data where these are significantly different.